\documentclass{ws-procs9x6}

\begin{document}

\title{Electroweak Measurements at CDF \footnote{\uppercase{O}n behalf
of the \uppercase{CDF} collaboration}}

\author{A.~Sidoti \footnote{\uppercase{W}ork supported by \uppercase{R}esearch
\uppercase{T}raining \uppercase{N}etwork of \uppercase{E.U.} ``\uppercase{P}robe of \uppercase{N}ew \uppercase{P}hysics''
\uppercase{HPRN-CT}-2002-00292 \uppercase{C}ontract}}

\address{Laboratoire de Physique Nucl\`eaire et de Hautes Energies \\
Universit\`e ``Pierre et Marie Curie'' (Paris VI) \\
4, Place Jussieu \\
75252 Paris Cedex 05, France}

\maketitle

\abstracts{
We present some recent measurements on electroweak physics
using data collected by the CDF experiment at the Tevatron proton
anti-proton collider ($\sqrt{s}=1.96\mathrm{TeV}$)  
at Fermilab (Batavia, Ill, USA).
}

\section{Introduction}
The CDF electroweak physics program is one of the key components of
the RunII. Electroweak measurements  are complementary to those
performed at $e^+e^-$ machines (LEP and SLD). The
former can produce a larger number of W bosons and can produce
$Z/\gamma^\star$ at higher invariant mass with respect of the latters.

We will review CDF electroweak physics measurements using data
collected  from February 2002. The integrated luminosity  of data
ranges from 64 
pb$^{-1}$ to $\sim$200 pb$^{-1}$ depending on the measurement. 

\section{W and Z inclusive cross section measurements}
\begin{figure}[ht]
\begin{minipage}{.49\linewidth}
\centerline{\epsfxsize=1.8in\epsfbox{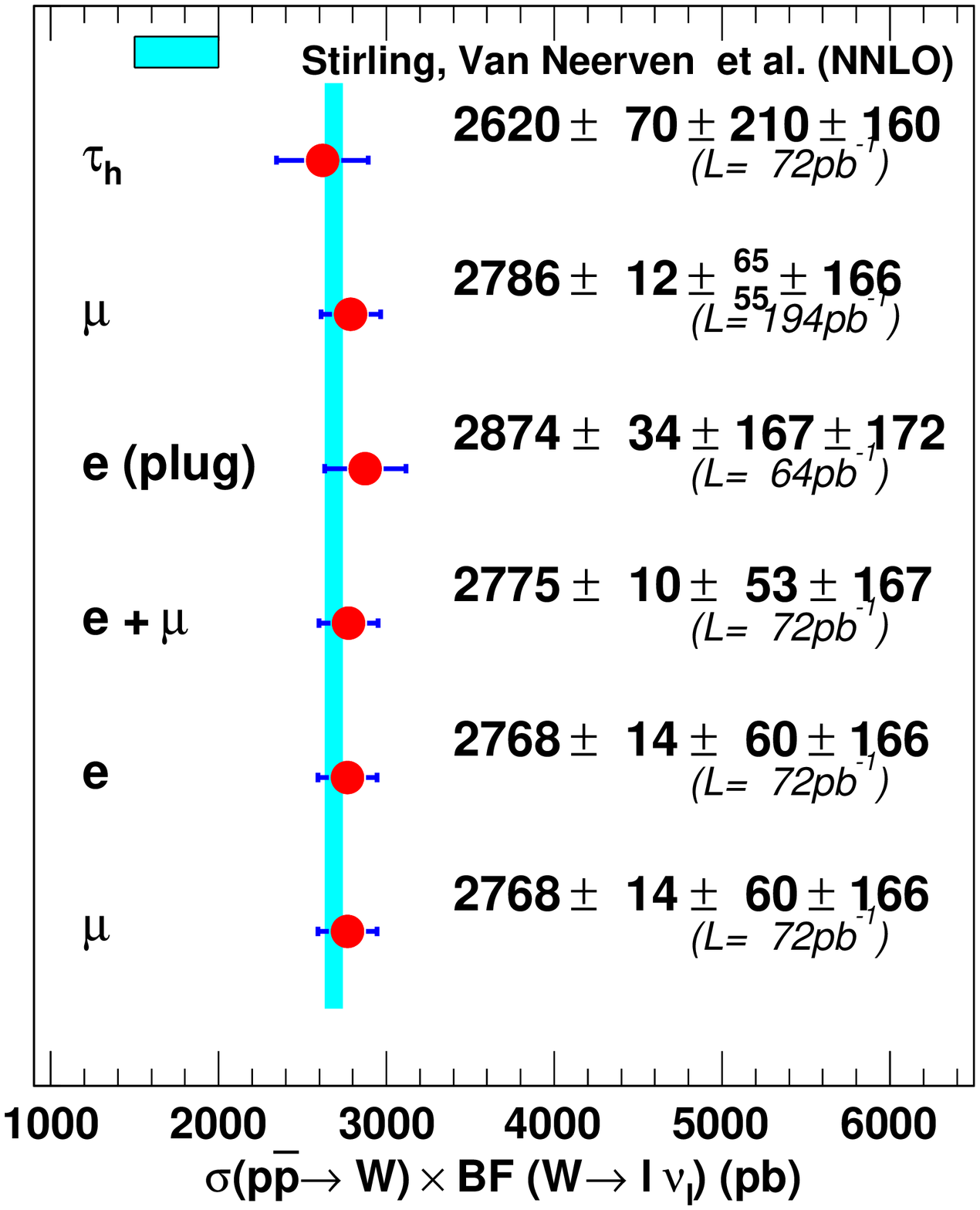}}   
\end{minipage}
\begin{minipage}{.49\linewidth}
\centerline{\epsfxsize=1.8in\epsfbox{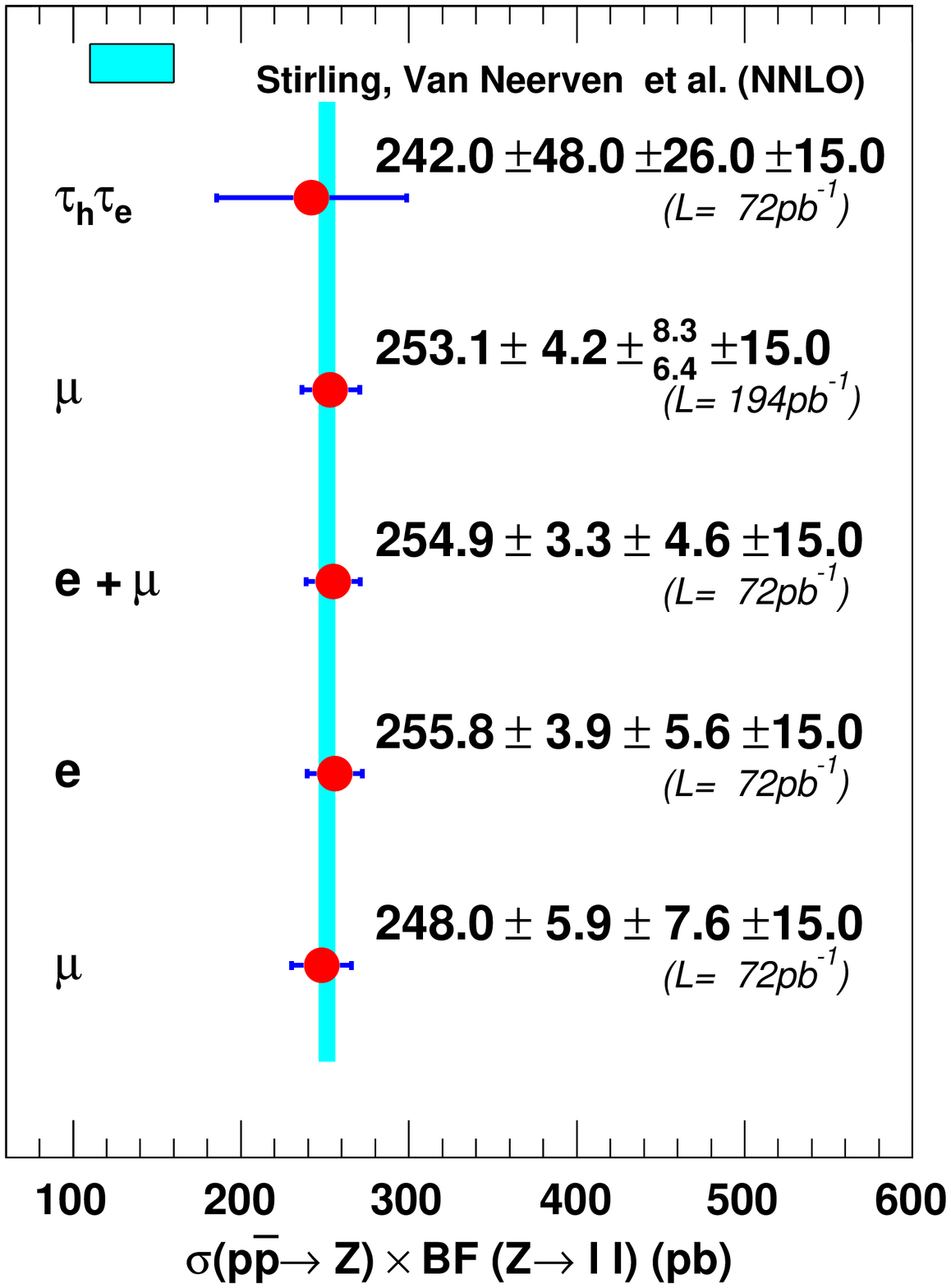}}  
\end{minipage}
\caption{$\sigma(p\bar{p}\rightarrow W) \times BF(W\rightarrow \ell
\nu_{\ell})$  (left) and 
$\sigma(p\bar{p}\rightarrow Z) \times BF(Z\rightarrow \ell
\ell)$  (right) measured at CDF. The blue band indicates the
theoretical (NNLO) predictions. \label{fig:wz_xsec}}
\end{figure}

Due to the high branching ratios and clean signature W and Z bosons are
identified through their leptonic decays. Inclusive cross sections of
both W and Z 
have been measured using all the three leptons: electrons, muons and
{\it taus} and using all the available subdetectors of CDF extending
in particular the geometric acceptance at high pseudorapidity
$\eta$\footnote{$\eta$ is related to the azimuthal angle through the
relation $\eta = -\log \tan (\theta/2)$}. All the measurements
performed are shown in Fig.\ref{fig:wz_xsec} and are in agreement with
the theoretical predictions 
\cite{wz_xsec_theo}.

\vspace{-0.7cm}

A stringent test of the Standard Model can be performed by evaluating
the ratio $R$ of the W and Z boson cross sections:
$$
R = \frac{\sigma(p\bar{p}\rightarrow W)\times\mathrm{BF}(W\rightarrow
\ell \nu_{\ell})}
{\sigma(p\bar{p}\rightarrow Z)\times\mathrm{BF}(Z\rightarrow
\ell \ell)}
$$
that can be written as 
$$
R = \frac{\sigma(p\bar{p}\rightarrow W)}{\sigma(p\bar{p}\rightarrow
Z)} 
\times \frac{\Gamma(Z)}{\Gamma(Z\rightarrow \ell \ell)} \times 
\frac{\Gamma(W\rightarrow \nu \nu_{\ell})} {\Gamma(W)}.
$$ 
Inserting the theoretical values of the total cross sections and of
the partial width $W\rightarrow \ell \nu_{\ell}$ and  the experimental
partial and total widths of the Z boson from LEP, it is possibile to
extract indirectly the total width of the W boson.
The measured value by CDF is:
\begin{eqnarray*}
\Gamma(W) =& 2079\pm41 \;\mathrm{MeV} \;\;\; \mathrm{CDF}\; e+\mu
\;\mathrm{channel}\;\int\mathcal{L}dt = 72\;\mathrm{pb}^{-1}  \\
\Gamma(W) =& 2056\pm44 \;\mathrm{MeV} \;\;\; \mathrm{CDF}\; \mu
\;\mathrm{channel}\;\int\mathcal{L}dt = 194\;\mathrm{pb}^{-1} 
\end{eqnarray*}
well in agreement with both the PDG world average\cite{pdg} and the
theoretical predictions:
$$
\Gamma(W)_{\mathrm{PDG}} = 2118\pm41 \;\mathrm{MeV}  \;\;\;\;
\Gamma(W)_{\mathrm{Th.}} = 2092.1 \pm 2.5\; \mathrm{MeV}
$$


\section{W Boson Asymmetries}

\begin{figure}[ht]
\centerline{\epsfxsize=3.7in\epsfbox{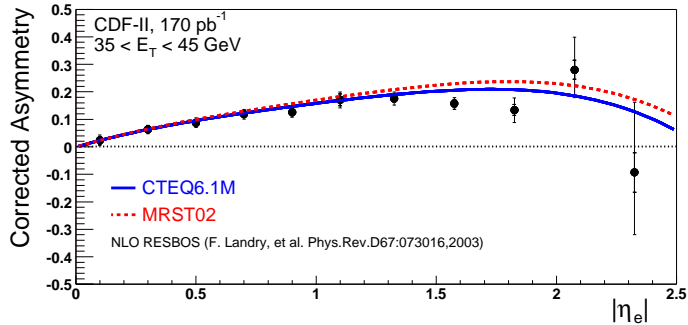}}   
\caption{W charge asymmetry as a function of the electron
pseudorapidity, with 
$\;\;\;\;\;\;\;35<E_{T}<45$\,GeV. Comparison with expectations
from different pdf are shown.
\label{fig:w_charge}}
\end{figure}

A precise charge asymmetry as a function of rapidity of the W boson
provides constraints on the 
parton fluxes of the incoming protons and therefore provides a better
determination of the parton distribution functions (pdf). The leptonic
decay of the W boson makes difficult to measure directly the W rapidity
itself. Instead, the electron asymmetry is measured that is the
convolution of the W production charge asymmetry and the V-A asymmetry
from the W decay.
The lepton charge aymmetry is defined as:
$$
A(\eta_{\ell}) =
\frac{d\sigma_{+}/d\eta_{\ell}-d\sigma_{-}/d\eta_{\ell}}
{d\sigma_{+}/d\eta_{\ell}+d\sigma_{-}/d\eta_{\ell}}
$$
where $\eta_{\ell}$ is the pseudorapidity of the lepton. CDF has
measured the W charge asymmetry in the electron channel. The result as
a function of the electron rapidity is shown in
Fig.\ref{fig:w_charge}. The measurement has been corrected for the
effect of charge mis-identification and  background contributions
both dependent on the pseudorapidity. An additional selection
$35<E_{T}<45$ GeV is 
applied on the electron energy to increase the sensitivity
to different pdf.  In fact the direction of an  electron with higher
energies is closer to the direction of emission of the W boson
enhancing therefore the W production asymmetry\cite{w_charge_asym}
\vspace{-0.7cm}

\hspace{-0.5cm}

\section{Di-Boson Production}
Di-boson processes probe electroweak gauge bosons interactions as well
as sources of physics beyond Standard Model. 
\subsection{$W\gamma$ and $Z\gamma$ production}
\begin{figure}[ht]
\begin{minipage}{.49\linewidth}
\centerline{\epsfxsize=1.9in\epsfbox{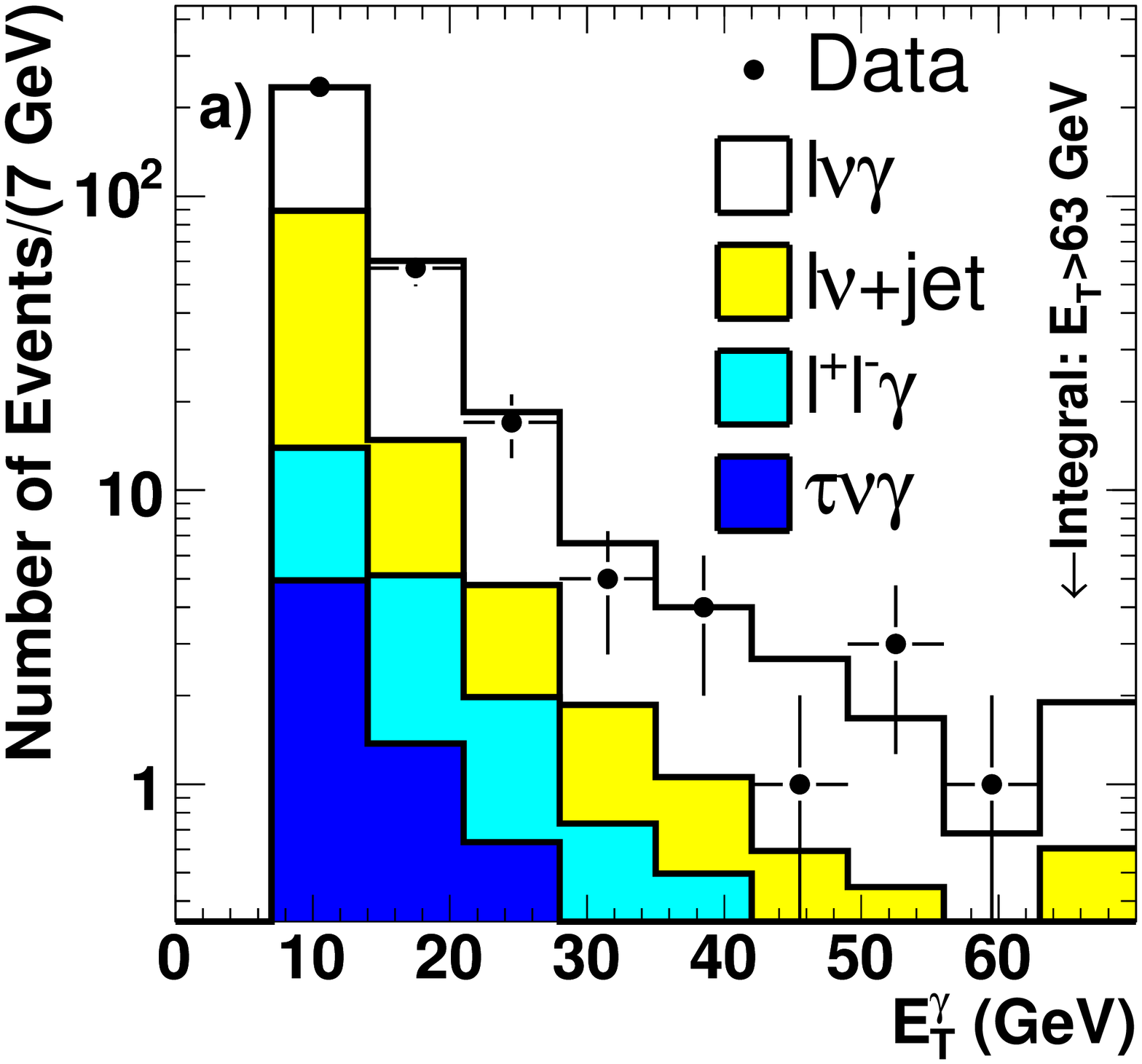}}   
\end{minipage}
\begin{minipage}{.49\linewidth}
\centerline{\epsfxsize=1.9in\epsfbox{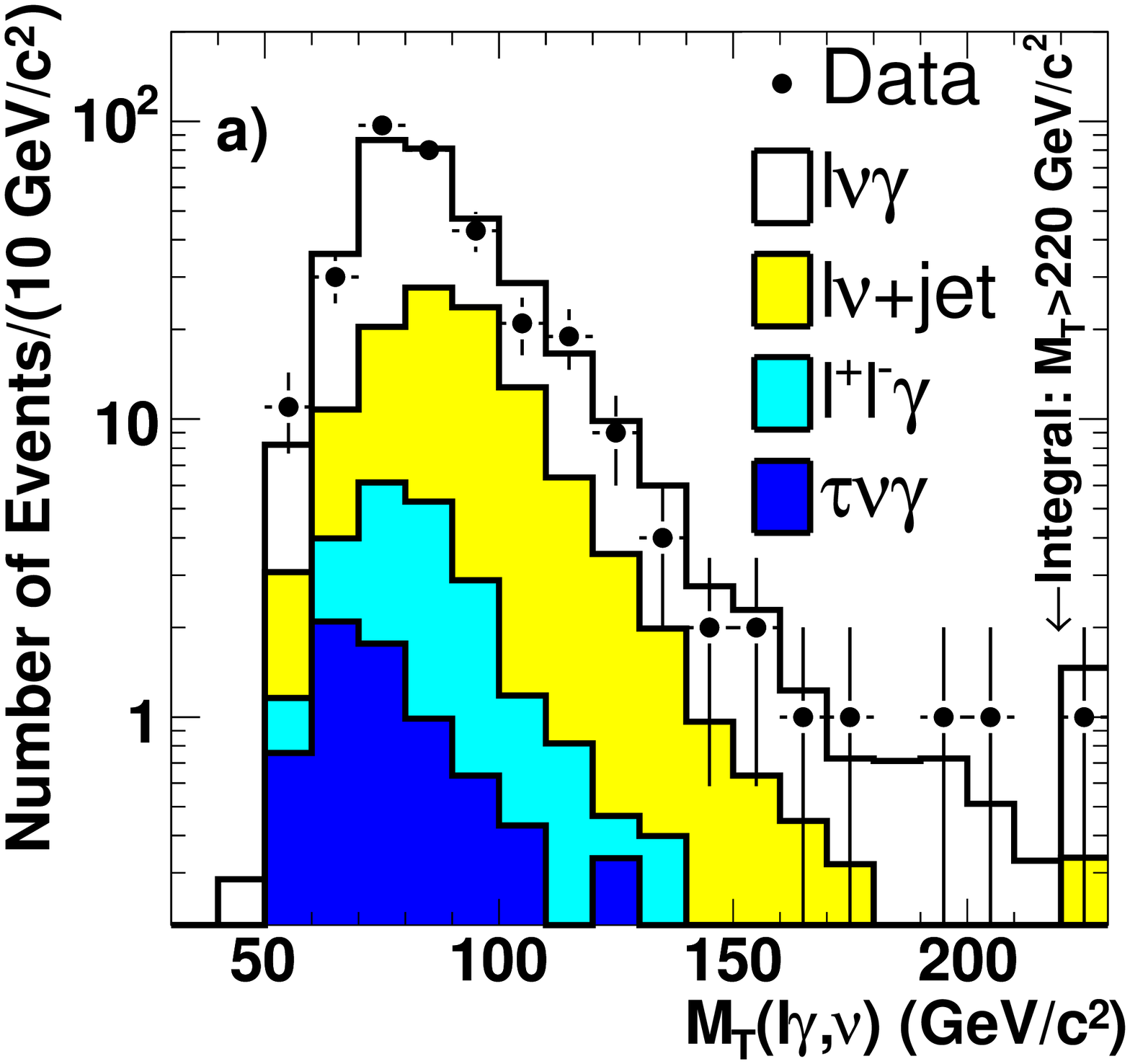}}  
\end{minipage}
\caption{Photon $E_{T}$ (left) and Cluster transverse mass (right) for
$W\gamma\rightarrow \ell \nu \gamma$ 
shown for observed events together with signal and background
expectations. Anomalous gauge coupling will increase the high energy
tail of the photon $E_{T}$ distribution. \label{fig:wgamma}} 
\end{figure}

When one of the bosons is
a photon the main backgounds are $\pi_{0}$ and jets faking a
photon. To reduce these, CDF exploits the high spatial resolution of the
electromagnetic calorimeter provided by the {\it showermax}
detector\cite{showermax} 
and a track isolation criterium. 
The $W\gamma \rightarrow \ell \nu_{\ell} \gamma$ process has been
studied by CDF in the electron and muon channel. Candidate events are
selected requiring a  high-$P_{T}$
momentum lepton with large missing transverse energy and $M_{T}(\ell,\nu)<120$\,GeV/$c^2$. The presence of
a  photon with $E_{T}>7$ GeV well separated from the lepton
($\Delta R(\ell,\gamma)>0.7$\footnote{$R=\sqrt{\Delta\phi^2+\Delta\eta^2}$}) is
required. The cross section times branching fraction measured is:
$$
\sigma \times \mathrm{BF}(W\gamma \rightarrow \ell \nu_\ell \gamma) =
18.1 \pm 1.6\mathrm{(stat.)}\pm2.4\mathrm{(syst.)}\pm
1.2\mathrm{(lum.)} \mathrm{pb}^{-1}
$$
The predicted cross section including the photon
acceptance is \cite{bar}: 
$$
\sigma \times \mathrm{BF}(W\gamma \rightarrow \ell \nu_\ell
\gamma)_{\mathrm{Th.}} =
19.3 \pm 1.4 \; \mathrm{pb}
$$
The photon transverse energy is shown in
Fig.\ref{fig:wgamma}.  

$Z\gamma \rightarrow \ell \ell \gamma $ candidates are selected requiring two oppositely charged high
momentum leptons with a dilepton invariant mass such that
$M_{\ell,\ell}>40$ GeV/$c^2$. The photon selection is the same of the $W\gamma
\rightarrow \ell \nu_\ell \gamma$. 
The cross section times branching fraction measured is\cite{w_gamma}:
$$
\sigma \times \mathrm{BF}(Z\gamma \rightarrow \ell \ell \gamma) =
4.6 \pm 0.5\mathrm{(syst+stat)}\pm
0.3\mathrm{(lum.)} \mathrm{pb}^{-1}
$$
The predicted cross section including the photon
acceptance is \cite{bar}: 
$$
\sigma \times \mathrm{BF}(W\gamma \rightarrow \ell \nu_\ell
\gamma)_{\mathrm{Th.}} =
4.5 \pm 0.3
$$

\vspace{-0.7cm}


\subsection{WW Production}

WW production has been studied at CDF in the $WW\rightarrow \ell
\nu_\ell \ell' \nu_\ell'$ channel 
looking for two oppositely charged high $P_{T}$ leptons and large missing transverse
energy in the
final state. Using and integrated luminosity of 184 pb$^{-1}$ 17
candidate events have been found with an estimated 
background of $5.0^{+2.2}_{-0.8}$ events. The cross section 
measured is\cite{ww_prod}:
$$
\sigma(p\bar{p}\rightarrow WW) = 
14.6^{+5.8}_{-5.1}\mathrm{(stat)}^{+1.8}_{-3.0}\mathrm{(syst)} \pm
0.9\mathrm{(lum)} \mathrm{pb}
$$ 
well in agreement with the predicted cross section:
$$
\sigma(p\bar{p}\rightarrow WW)_{\mathrm{Th.}} =11.3\pm 1.3 
\mathrm{pb}
$$ 

\section{Conclusions}
CDF is producing interesting results in the electroweak sector.
Most of
them have a precision that is limited by statistics. Therefore we are
approaching a period where, with the available datasets, CDF can
produce electroweak physics measurements with the smallest single
experiment uncertainties. We are also refining the analysis using new
methods that will decrease the systematics uncertainties. 
So far, no SM deviations have been observed.

\section*{Acknowledgments}
It is a pleasure to thank all those at CDF and at Fermilab that are
working hard for the success of the RunII of the Tevatron. I would
also like to thank the organizers of this wonderful Lake Louise Winter
Institute.

\end{document}